\documentclass[aps,prb,preprint,showpacs,showkeys,floatfix]{revtex4}
\usepackage{graphicx,color}
\usepackage{microtype}
\usepackage{natbib}
\usepackage[utf8]{inputenc}
\usepackage{float}
\graphicspath{{figs/}}
\usepackage{amsmath}
\usepackage{url}
\makeatletter
\def\url@leostyle{%
  \@ifundefined{selectfont}{\def\UrlFont{\sf}}{\def\UrlFont{\small\ttfamily}}}
\makeatother
\usepackage{tabularx}
\usepackage[colorlinks,linkcolor=black,anchorcolor=black,citecolor=black, urlcolor=black]{hyperref}
\begin{document}

\title{Misfit strain-induced energy dissipation for graphene/MoS$_2$ heterostructure nanomechanical resonators}

\author{Ji-Dong He}
    \affiliation{Shanghai Institute of Applied Mathematics and Mechanics, 
    \\Shanghai Key Laboratory of Mechanics in Energy Engineering, 
    \\Shanghai University,~Shanghai~200072,~People's~Republic~of~China}
    
\author{Jin-Wu Jiang}
    \altaffiliation{Corresponding author: jwjiang5918@hotmail.com}
    \affiliation{Shanghai Institute of Applied Mathematics and Mechanics, 
    \\Shanghai Key Laboratory of Mechanics in Energy Engineering, 
    \\Shanghai University,~Shanghai~200072,~People's~Republic~of~China}

\date{\today}
\begin{abstract}

Misfit strain is inevitable in various heterostructures like the graphene/MoS$_2$ van der Waals heterostructure. Although the misfit strain effect on electronic and other physical properties have been well studied, it is still unclear how will the misfit strain affect the performance of the nanomechanical resonator based on the graphene/MoS$_2$ heterostructure. By performing molecular dynamics simulations, we disclose a misfit strain-induced decoupling phenomenon between the graphene layer and the MoS$_2$ layer during the resonant oscillation of the heterostructure. A direct relationship between the misfit strain and the decoupling mechanism is successfully established through the retraction force analysis. We further suggest to use the graphene/MoS$_2$/graphene sandwich heterostructure for the nanomechanical resonator application, which is able to prevent the misfit strain-related decoupling phenomenon. These results provide valuable information for the future application of the graphene/MoS$_2$ heterostructure in the nanomechanical resonator field.

\end{abstract}

% 78.20.Bh, Theory, models, and numerical simulation
% 63.22.-m, Phonons or vibrational states in low-dimensional structures and nanoscale materials
% 62.25.-g, Mechanical properties of nanoscale systems 
\maketitle
\pagebreak

\section{Introduction}
Van der Waals heterostructures have attracted intense research interest in recent years, owing to their multiple functional properties inherited from different constitute layers.\cite{Novoselov20162D, Geim2013Van} It was found that various van der Waals heterostructures can have good performance in the electronic devices, such as the MoS$_2$/MoSe$_2$ heterostructure~\cite{Terrones2013Novel, Kou2013Nanoscale, Kang2013Electronic} or the MoS$_2$/WS$_2$ heterostructure\cite{Tongay2014Tuning, Kun2015Electronic, Ko2013Electronic}. In 2013, Georgiou $\emph{et al.}$ reported a new generation of field-effect vertical tunnelling transistor based on the graphene/WS$_2$ heterostructures for flexible and transparent electronics.\cite{Thanasis2013Vertical} Guo $\emph{et al.}$ studied phosphorene and graphene heterostructures as negative electrode materials for rechargeable lithium batteries.\cite{Guo2015First}

Among other applications, few experiments have been carried out to investigate possible applications of the van der Waals heterostructure in the nanomechanical resonator field. Some researchers studied the mechanical properties of the multilayer graphene mechanical resonators coupled to superconducting cavities.\cite{Singh2014Optomechanical, Weber2016Force, Weber2014Coupling} She $\emph{et al.}$ studied the air damping effect on the multilayer MoS$_2$ nanomechanical resonator.\cite{She2016The} Some researchers examined the effect of the van der Waals interaction on multi-layered graphene mechanical resonators.\cite{He2005Resonance, Liu2011The, Arghavan2012Effects, Multilayer2009Youb} Ye $\emph{et al.}$ investigated the nanomechanical resonant oscillation for the graphene/MoS$_2$ (GM) heterostructure, and the Q-factor was measured to be 122.\cite{Ye2017Atomic}

As a characteristic feature for the van der Waals heterostructure, there is an inevitable misfit strain between the constituting atomic layers, due to different lattice constants for these different atomic layers.\cite{Jiang2014Mechanical, Lin2014Atomically, Li2014Structures} Lots of works have illustrated the importance of the misfit strain on the physical properties for the van der Waals heterostructure.\cite{S1997Misfit, Osbourn2003Strained, Morkoc1993Strained, Adams1992Semiconductor} However, the effect of the misfit strain on the nanomechanical resonator based on the van der Waals heterostructure is still unclear, which will be discussed in the present work.

In this letter, we study the namomechanical resonator based on the GM heterostructure. We find that the misfit strain is $3.1\%$ for the armchair GM heterostructure with length 86.8~{\AA}, due to different lattice constants for the graphene layer and the MoS$_2$ layer. As a direct result of the misfit strain, there is a decoupling phenomenon between the graphene and MoS$_2$ layers during the resonant oscillation of the GM heterostructure, which serves as a strong energy dissipation for the resonant oscillation. Based on the retraction force analysis, we explore the relationship between the misfit strain and the decoupling phenomenon for the GM heterostructure resonator.

\section{Structure and Simulation Details}

\begin{figure}[htpb]
  \begin{center}
    \scalebox{2}[2]{\includegraphics[width=4cm]{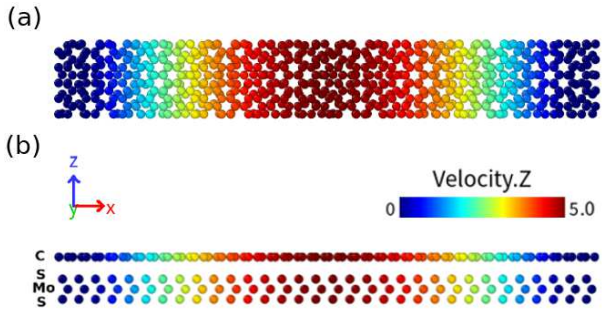}}
  \end{center}
  \caption{(Color online) Structure of the GM heterostructure of dimension $86.8\times 12.5$~{\AA}. (a) Top view. (b) Side view. The colorbar represents the velocity of each atom. The resonant oscillation is actuated by adding a sinuous distributed velocity to the system. }
  \label{fig_structure}
\end{figure} 

Figure.~\ref{fig_structure} illustrates one structure for the GM heterostructure of dimension $86.8\times 12.5$~{\AA} simulated in the present work. The fixed boundary condition is applied in the x-direction, while the periodic boundary condition is applied in the y-direction. The interactions in the graphene are described by the Brenner potential.~\cite{Donald2002A} The interactions in the MoS$_2$ are described by the Stillinger-Weber potential~\cite{Frank1985Computer} with parameters from recent works.~\cite{Zhou17Parameterization, Jiang2018Misfit} The van der Waals interaction between the graphene layer and the MoS$_2$ layer is described by the Lennard-Jones potential $E=4\epsilon[\left(\frac{\sigma}{r}\right)^{12}-\left(\frac{\sigma}{r}\right)^{6}]$  with parameters $\epsilon=0.00836$~eV and $\sigma=3.28$~{\AA}.\cite{JiangJW2015jap} 

The MD simulations are performed using the publicly available simulation code LAMMPS,\cite{PlimptonS,Lammps} while the OVITO package was used for visualization~\cite{Alexander2010Visualization}. The standard Newton equations of motion are integrated by using the velocity Verlet algorithm with a time step of 1 fs. There are three typical steps in the simulation of the resonant oscillation of the GM heterostructure. First, the system is thermalized to a constant temperature within the NVT (i.e. the particles number N, the volume V and the temperature T of the system are constant) ensemble. The constant temperature is maintained by the Nos\'e-Hoover thermostat.\cite{Nose,Hoover} Second, the resonant oscillation of the atomic layered material is actuated by adding a sinuous velocity distribution $v_z=v_0\sin({\pi}*x/L_x)$ as shown in Fig.~\ref{fig_structure}, where $v_0$ is the actuation velocity. Third, the system is allowed to oscillate within the NVE (i.e., the particles number N, the volume V, and the energy E of the system are constant) ensemble. The resonant oscillation energy from the third step is output to analyze the energy dissipation for the resonant oscillation of the system.

\section{Results and Discussions}

\begin{figure*}[htpb]
  \begin{center}
\scalebox{1.7}[1.7]{\includegraphics[width=9.5cm]{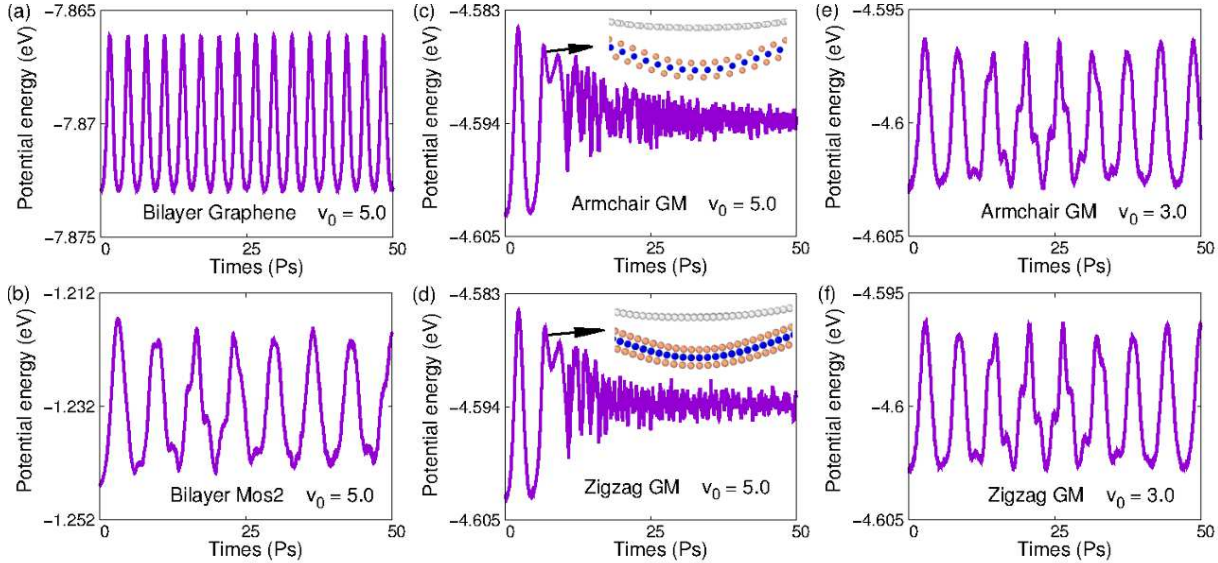}}
  \end{center}
  \caption{(Color online) The time history of the potential energy (per atom) for (a) bilayer graphene, (b) bilayer MoS$_2$, (c) GM heterostructure in the armchair direction, and (d) GM heterostructure in the zigzag direction with actuation velocity $v_0=5.0$~{\AA}/ps. The temperature is 4.2~K. Insets in (c) and (d) are MD snapshots for the central region at 7.0~{ps}. The time history of the potential energy (per atom) for the GM heterostructure along (e) the armchair direction and (f) the zigzag direction with actuation velocity $v_0=3.0$~{\AA/ps} at 4.2~K.}
 \label{fig_debonding}
\end{figure*}

\begin{figure}[htpb]
  \begin{center}
\scalebox{1}[1]{\includegraphics[width=8cm]{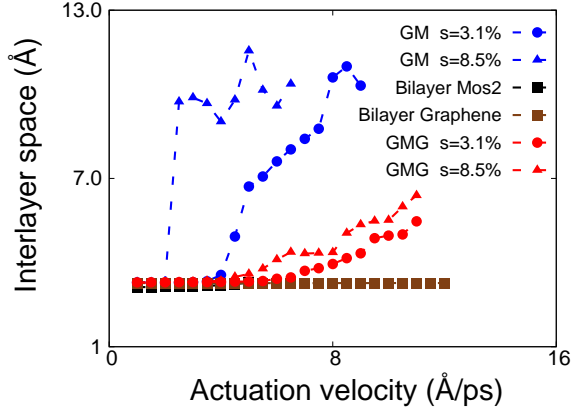}}
  \end{center}
  \caption{(Color online) The dependence of the interlayer space on the actuation velocity $v_0$ at 4.2~K for different structures.}
 \label{fig_distance}
\end{figure}

\begin{figure}[htpb]
  \begin{center}
\scalebox{1}[1]{\includegraphics[width=7cm]{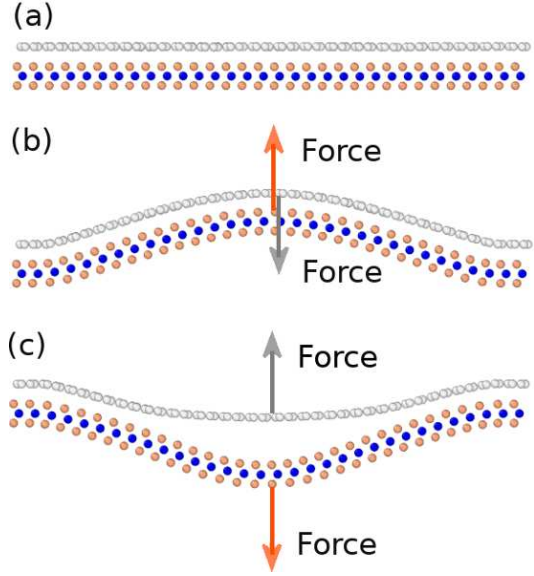}}
  \end{center}
  \caption{(Color online) The retraction force on the MoS$_2$ and graphene layers during the resonant oscillation. (a) The initial position. (b) The GM heterostructure moves in the upward direction. The forces on the MoS$_2$ and the graphene layers are in the opposite direction, resulting in the
compression of the GM heterostructure. (c) The GM heterostructure moves in the downward direction. The forces (in opposite direction) on the two atomic layers decouple the heterostructure.}
 \label{fig_force}
\end{figure}

\begin{figure}[htpb]
  \begin{center}
\scalebox{2}[2]{\includegraphics[width=4.5cm]{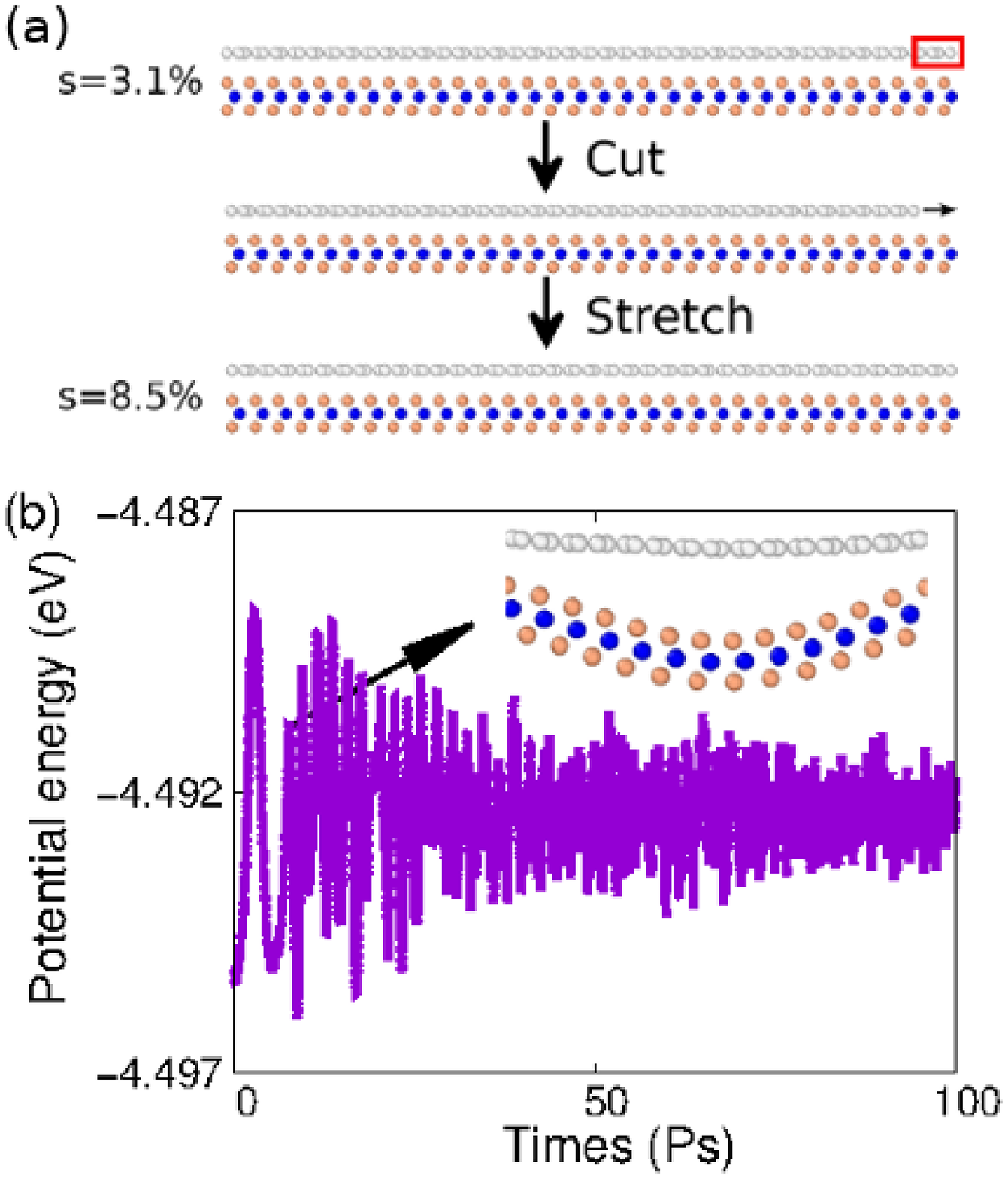}}
  \end{center}
  \caption{(Color online) The creation of the artificial misfit strain in the GM heterostructure. (a) The GM heterostructure of dimension $86.8\times 12.5$~{\AA} has a natural misfit strain $s=3.1\%$. One unit cell in the top graphene layer is removed, and the top graphene layer is slightly stretched. The misfit strain is thus increased to be $s=8.5\%$ manually. (b) The time history of the potential energy (per atom) for the GM heterostructure with artificial misfit strain $s=8.5\%$. The actuation velocity is $v_0=3.0$~{\AA/ps}, and the temperature is 4.2~K. Inset is a MD snapshot.}
 \label{fig_cut}
\end{figure}

Figures.~\ref{fig_debonding}~(a), (b), (c) and (d) compare the time history of the potential energy in the bilayer graphene, bilayer MoS$_2$, and the GM heterostructure. The actuation velocity is $v_0=5.0$~{\AA/ps} here. The dimensions for all structures are $86.8\times 12.5$~{\AA} for the armchair systems, and $87.5\times 21.7$~{\AA} for the zigzag systems. There is only weak energy dissipation in the bilayer graphene and bilayer MoS$_2$, as shown by panels (a) and (b). However, Figs.~\ref{fig_debonding}~(c) and (d) illustrate that the oscillation energy decays very fast in both armchair and zigzag GM heterostructure; i.e., there is strong energy dissipation in the GM heterostructure. The insets in Figs.~\ref{fig_debonding}~(c) and (d) disclose that there is a decoupling phenomenon between the top graphene layer and the bottom MoS$_2$ layer, which shall be responsible for the strong energy dissipation in the GM heterostructure resonator.

The decoupling phenomenon is in close relation to the actuation velocity $v_0$. Figs.~\ref{fig_debonding}~(e) and (f) show that there is no obvious energy dissipation in the GM heterostructure resonator for a smaller actuation velocity of $v_0=3.0$~{\AA/ps}, as there is no decoupling phenomenon. A quantitative relation between the decoupling phenomenon and the actuation velocity is summarized in Fig.~\ref{fig_distance}. The decoupling phenomenon is represented by the variation in the interlayer space between the two atomic layers in the central region. A sudden increase in the interlayer space reflects the presence of the decoupling phenomenon. We find that there is no decoupling phenomenon in the bilayer graphene and bilayer MoS$_2$ system for any large actuation velocity $v_0$. Note that the bilayer MoS$_2$ structure will be fractured during the resonant oscillation actuated by large $v_0$ above 5.0~{\AA/ps}. For armchair GM heterostructure, there is a critical value for the actuation velocity, above which the decoupling phenomenon takes place and the energy dissipation is very large accordingly. The critical actuation velocity is about 4.0~{\AA/ps} for the armchair GM heterostructure resonator. Zigzag GM heterostructure have similar situation as the armchair GM heterostructure.

To explore the underlying mechanism for the decoupling phenomenon and thus the strong energy dissipation, we notice that there is inevitable misfit strain between the graphene and the MoS$_2$ layers in the GM heterostructure, due to the difference in the lattice constants for these two atomic layers. The misfit strain~\cite{S1997Misfit} is obtained by
\begin{eqnarray}
s=|(l_{\rm top}-l_{\rm bot})/l_{\rm bot}|,
\label{eq_s}
\end{eqnarray}
where $l_{\rm top}$ and $l_{\rm bot}$ are the original length of the top and the bottom atomic layers. For bilayer graphene and bilayer MoS$_2$, the misfit strain is zero. The misfit strain is $s=3.1\%$ for the armchair GM heterostructure of length 86.8~{\AA}. As a result of this misfit strain, the top graphene layer is stretched by $0.7\%$, while the bottom MoS$_2$ layer is compressed by $-2.2\%$ in the relaxed configuration of the armchair GM heterostructure.

It is important that the misfit strain always has opposite sign for the graphene layer and MoS$_2$ layer in the armchair GM heterostructure, which results in opposite retraction force on the graphene layer and the MoS$_2$ layer during the oscillation process as shown in Fig.~\ref{fig_force}, where atomic layers just oscillate slightly away from their initial horizontal position. Fig.~\ref{fig_force}~(b) shows that the retraction force (denoted by arrows) on the top graphene layer is in the -$\hat{e}_z$ direction, because the potential energy within the graphene (stretched by the tensile misfit strain) is increased when the graphene layer is moving away from the horizontal position. However, the retraction force on the bottom MoS$_2$ layer is along the +$\hat{e}_z$ direction, since the potential energy of the MoS$_2$ layer (compressed by the compressive misfit strain) is lowered when it moves slightly away from the horizontal position. The overall effect of the retraction force is to compress the armchair GM heterostructure along the z-direction (out-of-plane direction) when the structure is above the horizontal direction as shown in Fig.~\ref{fig_force}~(b), so the decoupling phenomenon shall not occur in this region.

With similar analysis, we obtain the retraction forces on the graphene layer and the MoS$_2$ layer when the structure is in the lower region as shown in Fig.~\ref{fig_force}~(c). We find that the overall effect of the retraction force is to decouple the graphene layer and the MoS$_2$ layer. As a result, the decoupling phenomenon takes place at some point when the armchair GM heterostructure is in the lower region during the oscillation process. This is verified by our MD simulations, as all decoupling phenomena occur when the armchair GM heterostructures are in the lower region.

The effect of the misfit strain can be further enhanced by the difference in the resonant frequency of the graphene layer and the MoS$_2$ layer. More specifically, the resonant frequency for the graphene is 57.98~GHz, which is larger than the value of 41.20~GHz for the MoS$_2$ with similar length around 86.8~{\AA}. As a result, the misfit strain-induced compressive effect will be enhanced by the frequency difference in Fig.~\ref{fig_force}~(b). Similarly, the misfit strain-induced decoupling effect will also be enhanced by the frequency difference in Fig.~\ref{fig_force}~(c).

To further verify the relation between the misfit strain and the decoupling phenomenon, we investigate the armchair GM heterostructure with artificially engineered misfit strain as shown in Fig.~\ref{fig_cut}~(a). The original natural misfit strain is $s=3.1\%$. We cut one unit cell away from the top graphene layer, which is then slightly stretched. The misfit strain is thus artificially increased from $s=3.1\%$ to $s=8.5\%$. This GM heterostructure is relaxed, and the resultant strain is $s=2.3\%$ for the top graphene layer and $s=-6.1\%$ for the bottom MoS$_2$ layer. Fig.~\ref{fig_cut}~(b) shows that the decoupling phenomenon occurs even though the actuation velocity is a small value of $v_0=3.0$~{\AA/ps}. Note that there is no decoupling phenomenon for the armchair GM heterostructure (with natural misfit strain $s=3.1\%$) actuated with $v_0=3.0$~{\AA/ps} as shown in Fig.~\ref{fig_debonding}~(e). This result further indicates that the decoupling phenomenon is in close relation to the misfit strain.

In the above, we have established that the misfit strain in the GM heterostructure can cause the decoupling phenomenon of the MoS$_2$ and graphene layers. We find that such misfit strain-induced issue can be greatly improved by sandwiching the MoS$_2$ with another graphene layer, i.e., constructing the graphene/MoS$_2$/graphene (GMG) trilayer heterostructure. Fig.~\ref{fig_distance} shows that the critical value for the actuation velocity $v_0$ can be increased considerably in the GMG heterostructure. More specifically, the critical value for $v_0$ in the GMG heterostructure with natural misfit strain 3.1\% is about 7.0~{\AA/ps}, which is much larger than the value of 4.0~{\AA/ps} in the GM heterostructure. A larger critical value for $v_0$ indicates that the decoupling phenomenon is more difficult to occur in the GMG heterostructure. Similarly, the critical value for the actuation velocity in the GM heterostructure with artificial misfit strain 8.5\% is about 4.5~{\AA/ps}, which is much larger than the value of 2.5~{\AA/ps} in the GMG heterostructure.

\section{Conclusion}

To summarize, we have performed MD simulations to investigate the resonant oscillation of the GM heterostructure nanomechanical resonators. We find that the inevitable misfit strain in the GM van der Waals heterostructure leads to the decoupling between the graphene and the MoS$_2$ layer, resulting in the strong energy dissipation for the GM heterostructure resonator. Based on the analysis for the retraction force, we show that the decoupling phenomenon always takes place when the heterostructure is moving toward the MoS$_2$ side, which is verified by our MD simulations. The dependence of the decoupling phenomenon on the misfit strain can be further confirmed by manually increasing the misfit strain in the GM heterostructure. We also suggest to prevent this misfit strain-induced decoupling phenomenon by sandwiching the MoS$_2$ layer by two graphene layers, instead of utilizing the GM heterostructure.

\section*{Acknowledgment}

The work is supported by the Recruitment Program of Global Youth Experts of China, the National Natural Science Foundation of China (NSFC) under Grant Nos. 11504225 and 11822206, and the Innovation Program of Shanghai Municipal Education Commission under Grant No. 2017-01-07-00-09-E00019.

\bibliographystyle{apsrev}
\bibliography{bibtex}
\end{document}